\documentclass[preprints,article,accept,oneauthor]{Definitions/mdpi} 

\firstpage{1} 
\pubvolume{1}
\issuenum{1}
\articlenumber{0}
\pubyear{2025}
\copyrightyear{2025}
\datereceived{ } 
\daterevised{ }
\dateaccepted{ } 
\datepublished{ } 
\hreflink{https://doi.org/}

\usepackage{graphicx,xcolor}
\usepackage{extarrows}
\newcommand{\myvec}[1]{\mbox{\boldmath $#1$}}
\definecolor{daidai}{rgb}{1,0.5,0}
\definecolor{usucha}{rgb}{0.75,0.375,0}
\definecolor{kogecha}{rgb}{0.5,0.25,0}

\Title{Coupled Map Lattice for Astronomical Object Formation: A Scenario for Evolution from Star to Disk, Arms, and Companions}

\TitleCitation{Coupled Map Lattice for Astronomical Object Formation: A Scenario for Evolution from Star to Disk, Arms, and Companions}

\Author{Erika Nozawa $^{1,}$*}

\AuthorNames{Erika Nozawa}

\AuthorCitation{Nozawa, E.}

\address[1]{\vspace{-0.4cm}
$^{1}$ \quad Graduate School of Organic Materials Science, Yamagata University, 4-3-16 Johnan, Yonezawa, Yamagata 992-8510, Japan\\
* \hspace{0.25cm} Correspondence: papers@e-rika.net}

\corres{Correspondence: papers@e-rika.net}

\abstract{We present a new dynamic formation model of a star, a disk, arms, and companions using a coupled map lattice (CML), a complex systems approach. This CML simulates the viscoelastic and chaotic dynamics and evolution of gas clumps containing a little dust with a minimal set of one Eulerian procedure for the flow formation of gas clumps due to gravitational interaction, and one Lagrangian procedure for the collision and mixture of gas clumps due to viscoelastic advection. Despite its simplicity, this CML successfully obtains four typical astronomical objects consistent with protoplanetary disk observations: a central star, Keplerian disk, spiral arms, and even stellar, substellar, and planetary companions. All these formation processes are truly dynamic, with the central star ``starring'' in them, and they are not based on the conventional disk gravitational instability but on the central star gravitational instability with high-dimensional chaotic gas ejection, namely the chaotic itinerancy. Of particular note is the process in which diverse companions are formed due to the rapid density increase caused by the intersection of spiral arms. This suggests a novel companion formation scenario that should be called ``arm-crossing companion formation'' with a view to planet formation, which may overcome the radial drift barrier and angular momentum problem.}


\begin{document}
\hypersetup{hypertexnames=false}
\hypersetup{linkcolor=usucha, citecolor=daidai, urlcolor=kogecha}

\section{Introduction}
\label{Introduction}

Coupled map lattice (CML) is a spatio-temporal dynamical system with discrete space and time and continuous state variables, and it is known as a powerful complex systems approach \cite{Kaneko,Kanekos,Kanekob}. CMLs well reproduce experimental observations in spatially extended dynamic phenomena due to their flexible construction with parameterized nonlinear maps (i.e., the Eulerian and Lagrangian procedures) \cite{Yanagitab,Yanagitac,Yanagitad,Nishimori}. Furthermore, CMLs offer new and insightful perspectives underlying such phenomena due to their fast simulations, including wide-range parameter searches that are fully executable at the personal computer level. In fact, in the previous papers \cite{Nozawa,Nozawa2,Nozawa3}, we proposed a CML for the spiral arm formation of gas clumps, such as seen in spiral galaxies and protoplanetary disks, and obtained the following simulation results that not only agreed with conventional theories and observations but also provided new suggestions. First, the simulated spiral arms are conventional pattern arms similar to density waves \cite{Lin-Shu,Kuno,Kuno2,Laura} and are newly interpreted as astronomical traffic jams formed by jammed Keplerian gas \cite{Nozawa,Nozawa2}. Second, the spiral arms are transient, as suggested in the observation \cite{Meidt}, and are consistently explained as they disappear due to the gas flow rate difference ``light-in and heavy-out'' between a light gas inflow into and heavy outflow from the jam \cite{Nozawa2}. Finally, the remaining lifetime of the spiral arm in M51 is reasonably predicted as about 150 million years (Myr) using an approximate formula applicable to actual galaxies directly derived from the ``light-in and heavy-out'' approach \cite{Nozawa2,Nozawa3}.

In protoplanetary systems, the formation of a central star is followed by the formation of a Keplerian disk, spiral arms, and diverse companions, including stars, brown dwarfs, and planets. Our previously proposed CML \cite{Nozawa,Nozawa2,Nozawa3}, which exhibits astronomical chaotic itinerancy (CI) \cite{Konishi}, succeeded in forming a central star and the two additional types of objects of a Keplerian disk and spiral arms formed by high-dimensional chaotic gas ejection \cite{Yanagitaa,Martin} with the expansion and contraction of the central star, for a total of three types of objects. However, due to the low gas ejection masses, it could not form the last but most important type of objects: stellar, substellar, and planetary companions. The reason for this should be that the dust effect on the gas clump motion was not taken into account in this previous CML. Dust is essential in the formation of the companions (especially planets), as is discussed on the issues in protoplanetary systems, such as the ``radial drift barrier'' \cite{Kataoka,Marel} and ``angular momentum transport''  \cite{Kalyaan,Koga}.

This paper reports on a CML for astronomical object formation (hereinafter referred to as ``Astro CML'') that consistently covers a series of formation processes of four types of astronomical objects, whose existence has been confirmed by observations of protoplanetary systems, including a central star, a Keplerian disk, spiral arms, and various stellar, substellar, and planetary companions. The Astro CML focuses on the viscoelastic and chaotic dynamics and evolution of discrete gas clumps containing small amounts of dust rather than on the dynamics and evolution of dust in a general continuous gas medium. A sequence of evolution of gas clumps is simulated by a minimal set of one Eulerian procedure for the flow formation of gas clumps due to gravitational interaction, and one Lagrangian procedure for the collision and mixture (i.e., deformation) of gas clumps due to viscoelastic advection. Note that the Astro CML was proposed in the author's presentation in JPS \cite{Nozawa4}, was awarded by the JPS Student Presentation Award \cite{Nozawa5}, and was summarized in the doctoral dissertation \cite{Nozawa6}, but is being presented for the first time in academic journals.

Despite its simple model concept, the Astro CML succeeds in reproducing all four types of astronomical objects in protoplanetary systems. All of each formation process are truly dynamic, with the central star ``starring'' in them, and they are not based on the conventional disk gravitational instability but on the star gravitational instability. The central star is stably formed through rapid relaxation, even if it starts from a nonstationary initial state, and the Keplerian disk and spiral arms, including small and light ones in the earliest stage \cite{Okoda}, are formed due to high-dimensional chaotic gas ejection \cite{Yanagitaa}, i.e., astronomical CI \cite{Konishi}, from the pulsating central star under gravitational instability. Of particular note is the process of companion formation, in which companions in the planet or brown dwarf mass range are formed through the rapid density increase caused by the intersection of the old and new spiral arms due to the repetition of gas ejection from the central star, and the formed companions also form spiral arms while increasing their mass to promote subsequent companion formation along with the central star. This suggests a novel companion formation scenario that should be called ``arm-crossing companion formation'' with a view to planet formation, which may overcome the radial drift barrier and angular momentum problem.

A series of the aforementioned formation processes is classified into five distinctive evolutionary stages using the time series of the mass and angular momentum of the central star, disk, arm, and companion. (1) {\em Central star formation stage}: Gas clumps in a molecular cloud transport their mass and also angular momentum (including a small angular momentum generated by gas-dust viscoelastic interaction) to the central part of the molecular cloud (i.e., the central star), as they repeatedly fall and deform due to their gravitational interaction and viscoelastic advection. (2) {\em Disk and spiral arm formation stage}: The formed central star gradually decreases its mass while repeating high-dimensional chaotic gas ejections due to its gravitational instability, forming and growing a Keplerian disk and spiral arms while transporting mass and angular momentum to its surroundings. (3) {\em Disk and spiral arm maturation stage}: Once the Keplerian disk and spiral arms have grown sufficiently, gas ejection from the central star reaches a lull, and mature and balanced grand design spiral arms appear as the central star, disk, and arms hold constant mass and angular momentum. (4) {\em Companion formation stage}: The old spiral arm, which retains mass and angular momentum, is crossed by a new spiral arm that transports mass and angular momentum due to gas ejection from the central star, where a sudden increase in mass and angular momentum forms a companion in the planet or brown dwarf mass range without experiencing the radial drift, which grows while absorbing the mass and angular momentum of the matured disk. (5) {\em Companion maturation stage}: When the companion has grown sufficiently to lull the absorption of mass and angular momentum from the disk, a mature companion with constant mass and angular momentum also forms spiral arms and leads the subsequent companion formation along with the central star.

It should be emphasized that these evolutionary stages emerged spontaneously as a result of a series of evolutions in the collective behavior of gas clumps under consistent chaotic dynamics due solely to the gravitational interaction and viscoelastic advection of gas clumps containing small amounts of dust, which is an insightful scenario suggested by the Astro CML. Interestingly, the mass ratio of the formed companion to the central star is consistent with the most typically observed mass ratio \cite{Duquennoy,Chauvin}.

The present paper is organized as follows: Section~\ref{Model} constructs the Astro CML in accordance with the CML formalism of introducing a lattice, assigning field variables, formulating procedures, and defining time evolution. Section~\ref{Dynamic formation of a star, a disk, arms, and companions} investigates a series of formation processes from a central star to a Keplerian disk, spiral arms, and diverse companions, by classifying them into five distinctive evolutionary stages based on the simulation snapshots and the timeseries of mass and angular momentum. Section~\ref{Summary and discussion} contains a summary and discussion. Appendix~\ref{Viscoelastic advection} provides a derivation of viscoelastic advection under the concept of gas-clad dust particles.

\section{Model}
\label{Model}

Let us construct the Astro CML (a coupled map lattice for astronomical object formation) that consists of a minimal set of procedures: gravitational interaction and viscoelastic advection, acting on gas clumps. Sections \ref{Gas clumps on the lattice}-\ref{Dynamics and evolution of gas clumps} provide a step-by-step modeling of the chaotic dynamics that tracks the spontaneous evolution of gas clumps based on their viscoelastic behavior triggered by a little dust, according to the CML formalism \cite{Kaneko,Kanekos}.

\subsection{Gas clumps on the lattice}
\label{Gas clumps on the lattice}

The first construction step is to introduce coordinates to an object of interest using a suitable lattice. In the Astro CML, we consider cold dense gas clumps containing small amounts of dust in motion on a two-dimensional disk in three-dimensional space, such as are in protoplanetary systems \cite{Marel}. A two-dimensional square lattice is selected to introduce coordinates to this disk of gas clumps. The lattice points are denoted by $ij$ ($i=0,1,\cdots,N_{x}-1$ and $j=0,1,\cdots,N_{y}-1$), and their position vectors by $\myvec{r}_{ij}$ $=(i,j)=i\myvec{e}_{x}+j\myvec{e}_{y}$. Here, $\myvec{e}_{x}$ and $\myvec{e}_{y}$ are the unit vectors in the $x$- and $y$-axis directions, respectively.

\subsection{Field variables}
\label{Field variables}

The second construction step is to assign a set of field variables representing the state of the object at discrete time $t$ to the lattice. In the Astro CML, we consider the motion of gas clumps by gravity and viscoelastic forces, the most fundamental forces at work when gas clumps form astronomical objects, and introduce the gas clump mass $m_{ij}^{t}$ and the gas clump velocity $\myvec{v}_{ij}^{t}=v_{x\, ij}^{t}\myvec{e}_{x}+v_{y\, ij}^{t}\myvec{e}_{y}$ as field variables at lattice point $ij$. Here, the mass of gas is $(1-\alpha)m_{ij}^{t}$ and the mass of dust is $\alpha m_{ij}^{t}$ in a gas clump, using the mixing coefficient $\alpha$, which represents the mass fraction of dust in gas clumps. Note that the superscript $t$ denotes the index of time, not the exponent.

\begin{figure}[t]
\isPreprints{\centering}{}
\includegraphics[scale=0.5]{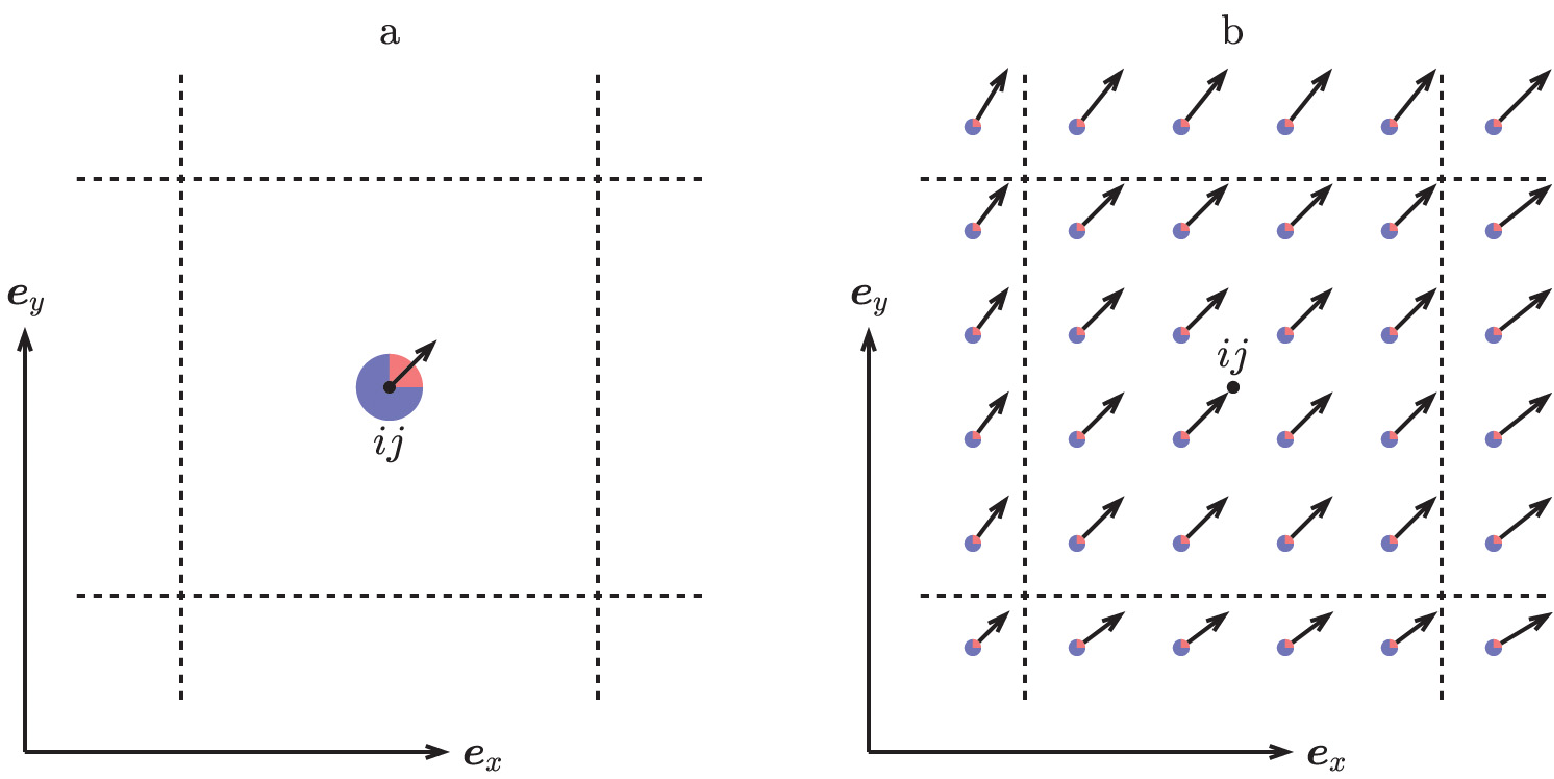}
\caption{Two complementary pictures of field variables, the gas clump mass $m_{ij}^{t}$ and velocity $\myvec{v}_{ij}^{t}$ at lattice point $ij$. (a) Lattice picture. The black dot denotes lattice point $ij$. The filled circle represents the mass of the gas clump, and the black arrow represents its velocity. In the circle, the blue part shows the gas mass $(1-\alpha)m_{ij}^{t}$, and the red part shows the dust mass $\alpha m_{ij}^{t}$ in the gas clump. (b) Particle picture. The black dotted square denotes the cell with size one, which is centered at lattice point $ij$. The set of the filled small circles represents the total mass of virtual particles made from gas and dust, and the set of the black arrows represents their flow. In each circle, the blue part shows the gas mass proportional to $(1-\alpha)m_{ij}^{t}$, and the red part shows the dust mass proportional to $\alpha m_{ij}^{t}$. The mixing coefficient (the dust mass fraction in gas clumps) is actually very small, but the figure is drawn with a value of 0.25 for visual clarity.\label{fig:field_variables.eps}}
\end{figure}

For flexible and proper formulation, there are two complementary pictures of the field variables in CMLs: the lattice picture and the particle picture, used in the construction of the Eulerian procedures and the Lagrangian procedures, respectively \cite{Nozawa} (which will be discussed shortly). Under the lattice picture (Figure~\ref{fig:field_variables.eps}a), the gas clump mass $m_{ij}^{t}$ and velocity $\myvec{v}_{ij}^{t}$ refer to the mass (the filled circle with blue gas and red dust parts in Figure~\ref{fig:field_variables.eps}a) and velocity (the black arrow in Figure~\ref{fig:field_variables.eps}a) of the gas clump at lattice point $ij$ (the black dot in Figure~\ref{fig:field_variables.eps}a), respectively. On the other hand, under the particle picture (Figure~\ref{fig:field_variables.eps}b), they refer to the total mass (the set of the filled small circles with blue gas and red dust parts in Figure~\ref{fig:field_variables.eps}b) and the flow (the set of the black arrows in Figure~\ref{fig:field_variables.eps}b) of virtual particles consisting of gas and dust in the square cell with size one (the black dotted square in Figure~\ref{fig:field_variables.eps}b) centered at lattice point $ij$, respectively. For simplicity, the virtual particles in each cell are treated as distributed uniformly and carried by the same flow. Note that virtual particles are not gas molecules or dust particles themselves.

\unskip

\subsection{Elementary processes and procedures}
\label{Elementary processes and procedures}

The third construction step is to formulate procedures describing the elementary processes of the state change of the object using relatively simple nonlinear maps acting on the field variables. In the Astro CML, we consider the most fundamental elementary processes of the dynamics and evolution of gas clumps with a little dust in astronomical object formation and formulate the gravitational interaction procedure $T_{g}$ and the viscoelastic advection procedure $T_{a}$.

As mentioned above, CMLs generally offer two types of procedures: the Eulerian and Lagrangian procedures \cite{Kaneko,Yanagitac}. The Eulerian procedures describe the change in field variables on each lattice point by local and global interactions, under the lattice picture, and the Lagrangian procedures describe the change in field variables in each cell along the flow of the virtual particles, under the particle picture. In Section \ref{Gravitational interaction procedure}, gravitational interaction is formulated as the Eulerian procedure $T_{g}$, and in Section \ref{Viscoelastic advection procedure}, viscoelastic advection is formulated as the Lagrangian procedure $T_{a}$, in the form of parameterized nonlinear maps, respectively.

\subsubsection{Gravitational interaction procedure}
\label{Gravitational interaction procedure}

In the gravitational interaction procedure $T_{g}$, the flow of gas and dust particles is formed in a gas clump at lattice point $ij$ due to the gravitational interaction (i.e., global interaction) from gas clumps at lattice points $kl$. The gas clump velocity $\myvec{v}_{ij}^{t}$ is changed to $\myvec{v}_{ij}^{*}$, where the symbol $*$ denotes an intermediate time between discrete times $t$ and $t+1$, by the following map:
\begin{align}
\label{eqn:Tg}
\myvec{v}_{ij}^{*}=\myvec{v}_{ij}^{t}+\tau_{g}\myvec{g}_{ij}^{t},
\end{align}
where $\tau_{g}$ is a time interval for the procedure $T_{g}$, and the gravitational field $\myvec{g}_{ij}^{t}=g_{x\, ij}^{t}\myvec{e}_{x}+g_{y\, ij}^{t}\myvec{e}_{y}$ at lattice point $ij$ is given by
\begin{align}
\label{eqn:g}
\myvec{g}_{ij}^{t}=\myvec{g}_{ij}\left(\left\{m_{kl}^{t}\right\}\right)=
-\gamma\sum_{k=0}^{N_{\mathstrut x}-1}\sum_{l=0}^{N_{\mathstrut y}-1}
\left(1-\delta_{ik}\delta_{jl}\right)\frac{m_{kl}^{t}}{\left|\myvec{r}_{ij}-\myvec{r}_{kl}\right|^{\mathstrut 2}}
\frac{\myvec{r}_{ij}-\myvec{r}_{kl}}{\left|\myvec{r}_{ij}-\myvec{r}_{kl}\right|^{\mathstrut}},
\end{align}
with the gravitational constant $\gamma$ and the Kronecker delta $\delta$. From Equations (\ref{eqn:Tg}) and (\ref{eqn:g}), the computational cost of the gravitational interaction procedure $T_{g}$ is $O((N_{x}N_{y})^{2})$.

\subsubsection{Viscoelastic advection procedure}
\label{Viscoelastic advection procedure}

In the viscoelastic advection procedure $T_{a}$, gas clumps at lattice points $kl$ are viscoelastically advected through a relaxation difference between light gas particles and heavy dust particles in the flows $\myvec{v}_{ij}^{*}$ resulting from the gravitational interaction (see Appendix \ref{Viscoelastic advection} for details), and collide and mix (i.e., deform) with a gas clump at lattice point $ij$ to form a new gas clump. The gas clump mass $m_{ij}^{t}$ and velocity $\myvec{v}_{ij}^{*}$ are changed to $m_{ij}^{t+1}$ and $\myvec{v}_{ij}^{t+1}$, respectively, by the following maps:
\begin{align}
\label{eqn:Ta_m}
m_{ij}^{t+1}&=\sum_{k=0}^{N_{\mathstrut x}-1}\sum_{l=0}^{N_{\mathstrut y}-1}w_{ijkl}^{*}m_{kl}^{t},
\\
\label{eqn:Ta_v}
\myvec{v}_{ij}^{t+1}&=\frac{1}{m_{ij}^{t+1}}\sum_{k=0}^{N_{\mathstrut x}-1}\sum_{l=0}^{N_{\mathstrut y}-1}
w_{ijkl}^{*}m_{kl}^{t}\myvec{v}_{kl}^{*},
\end{align}
where $w_{ijkl}^{*}$ is the weight of allocation from lattice point $kl$ to lattice point $ij$ (see \cite{Nozawa} for the derivation), and is given by
\begin{align}
\label{eqn:w_ijkl}
w_{ijkl}^{*}&=w_{ij}^{*}\left(\tilde{\myvec{r}}_{kl}^{*}\right)=w_{ij}^{*}\left(\tilde{k}^{*},\tilde{l}^{*}\right)
\nonumber\\
&=\left(\delta_{i\lfloor \tilde{k}^{*}\rfloor}\delta_{j\lfloor \tilde{l}^{*}\rfloor}
+\delta_{i\lfloor \tilde{k}^{*}\rfloor +1}\delta_{j\lfloor \tilde{l}^{*}\rfloor}\right.
\nonumber\\
&+\left.\delta_{i\lfloor \tilde{k}^{*}\rfloor +1}\delta_{j\lfloor \tilde{l}^{*}\rfloor +1}
+\delta_{i\lfloor \tilde{k}^{*}\rfloor}\delta_{j\lfloor \tilde{l}^{*}\rfloor +1}\right)
\left(1-\left|\tilde{k}^{*}-i\right|\right)\left(1-\left|\tilde{l}^{*}-j\right|\right),
\end{align}
with the floor function denoted by the symbol $\lfloor\,\rfloor$. Here, $\tilde{\myvec{r}}_{kl}^{*}=(\tilde{k}^{*},\tilde{l}^{*})$ is the position vector of the gas clump advected viscoelastically from lattice point $kl$, and it is given by
\begin{align}
\label{eqn:tr_kl}
\tilde{\myvec{r}}_{kl}^{*}=\tilde{\myvec{r}}_{kl}\left(\myvec{v}_{kl}^{*},\myvec{g}_{kl}^{t}\right)=\myvec{r}_{kl}+\left\{\left(1-\alpha\right)\myvec{v}_{kl}^{*}+\alpha\left(\myvec{v}_{kl}^{*}-\tau_{g}\myvec{g}_{kl}^{t}\right)\right\}\tau_{a},
\end{align}
with a time interval $\tau_{a}$ for the procedure $T_{a}$ (see Appendix \ref{Viscoelastic advection} for the derivation). Note that the computational cost of the viscoelastic advection procedure $T_{a}$ is only $O(N_{x}N_{y})$, rather than $O((N_{x}N_{y})^{2})$ expected from Equations (\ref{eqn:Ta_m}) and (\ref{eqn:Ta_v}). This is because the allocation weight matrix $w_{ijkl}^{*}$ is just a sparse matrix with four nonzero values, with respect to $ij$ when $kl$ is fixed (i.e., in each column of the matrix), as seen from Equation (\ref{eqn:w_ijkl}) (for details, see the appendix in \cite{Nozawa}).

\subsection{Dynamics and evolution of gas clumps}
\label{Dynamics and evolution of gas clumps}

The final construction step is to define the time evolution of the field variables describing the state change of the object by arranging a sequence of the procedures at each time step. In the Astro CML, we consider the order of dynamics and evolution of gas clumps in astronomical object formation and define the time evolution of the gas clump mass $m_{ij}^{t}$ and velocity $\myvec{v}_{ij}^{t}$ from discrete time $t$ to $t+1$ as follows:
\begin{align}
\label{eqn:dynamics}
\left\{
\begin{aligned}
&m_{ij}^{t}\rule{0pt}{1em} \\
&\myvec{v}_{ij}^{t}\rule{0pt}{1em} \\
\end{aligned}
\right\}
\xlongrightarrow[\substack{\text{Eulerian}\\ \text{procedure}}]{\substack{\text{\it Gravitational}\\ \text{\it interaction}\\ T_{g}\rule{0pt}{0.8em}}}
\left\{
\begin{aligned}
&m_{ij}^{t}\rule{0pt}{1em} \\
&\myvec{v}_{ij}^{*}\rule{0pt}{1em} \\
\end{aligned}
\right\}
\xlongrightarrow[\substack{\text{Lagrangian}\\ \text{procedure}}]{\substack{\text{\it Viscoelastic}\\ \text{\it advection}\\ T_{a\rule{0pt}{0.55em}}\rule{0pt}{0.8em}}}
\left\{
\begin{aligned}
&m_{ij}^{t+1}\rule{0pt}{1em} \\
&\myvec{v}_{ij}^{t+1}\rule{0pt}{1em} \\
\end{aligned}
\right\}.
\end{align}

The simulations were performed under the following conditions: The lattice size $N_{x}\times N_{y}$ is fixed as $50\times 50$, the initial gas clump mass $m_{ij}^{0}$ is given by a uniform random number within $[0,4/(N_{x}N_{y})]$, the initial gas clump velocity $\myvec{v}_{ij}^{0}$ is given by zero, and the open boundary conditions are assigned. The following parameter values were used: $\gamma=1$, $\tau_{g}=1$, and $\tau_{a}=1$, and the mixing coefficients $\alpha$ were set to $\alpha=0.05$ and $\alpha=0.06$, according to the typical mass fraction of dust to gas \cite{Miotello}.

On the time evolution of Equation (\ref{eqn:dynamics}), the total mass $\sum_{i,j}m_{ij}^{t}$, total momentum $\sum_{i,j}m_{ij}^{t}\myvec{v}_{ij}^{t}$, and total angular momentum $\sum_{i,j}\myvec{r}_{ij}\times m_{ij}^{t}\myvec{v}_{ij}^{t}$ of the system are conserved, except for a small angular momentum ($\propto\alpha$) arising from the elastic behavior of dust particles (for details, see Appendix \ref{Viscoelastic advection}). Note that, in the simulations, the total mass, total momentum, and total angular momentum are not completely conserved, since a small part of the gas and dust particles move out through the boundary of the finite lattice.

\begin{figure}[t]
\isPreprints{}{
\begin{adjustwidth}{-\extralength}{0cm}
\centering
}
\includegraphics[scale=0.93]{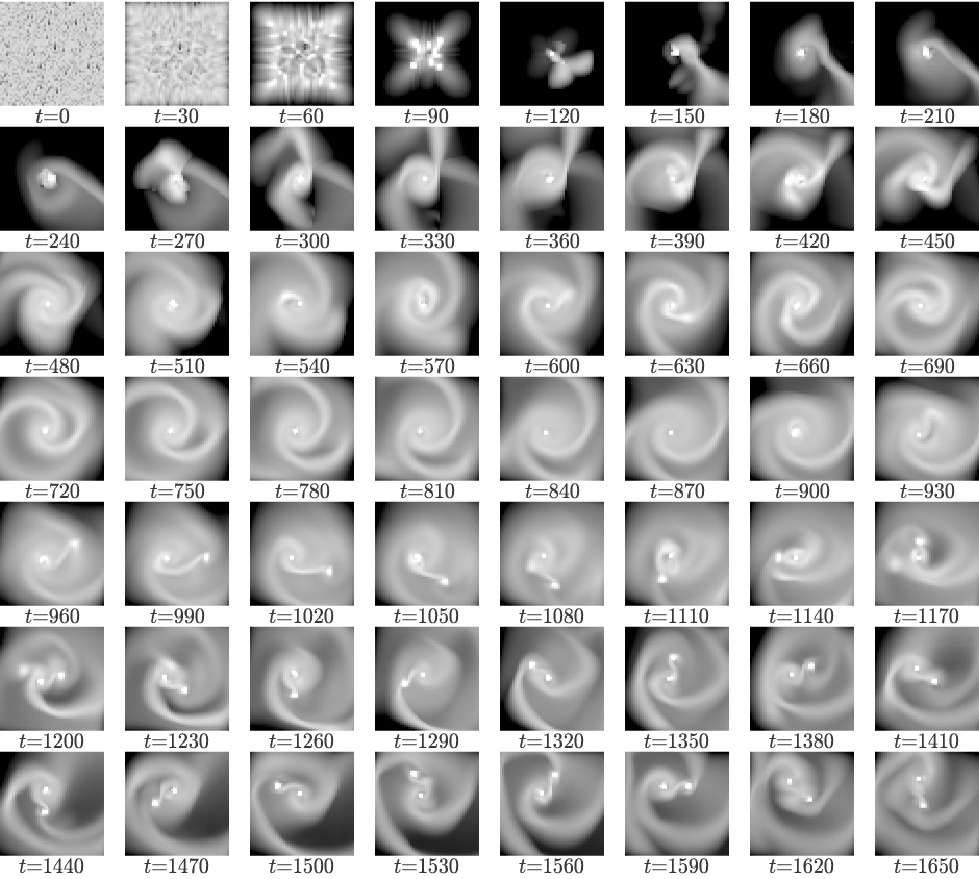}
\isPreprints{}{
\end{adjustwidth}
}
\caption{A series of formation processes at $\alpha=0.05$ from a central star to a Keplerian disk, spiral arms, and a companion. In each snapshot at $t=0$, $30$, $\cdots$, $1650$, the logarithm of gas clump masses, $\log_{10}m_{ij}^{t}$ ($i=0,1,\cdots,49$ and $j=0,1,\cdots,49$), is plotted in grayscale in the range of $1\times 10^{-8}$ to $1\times 10^{-2}$.\label{fig:ss_30stp_alpha_05(log).eps}}
\end{figure}

\begin{figure}[t]
\isPreprints{\centering}{}
\includegraphics[scale=0.5]{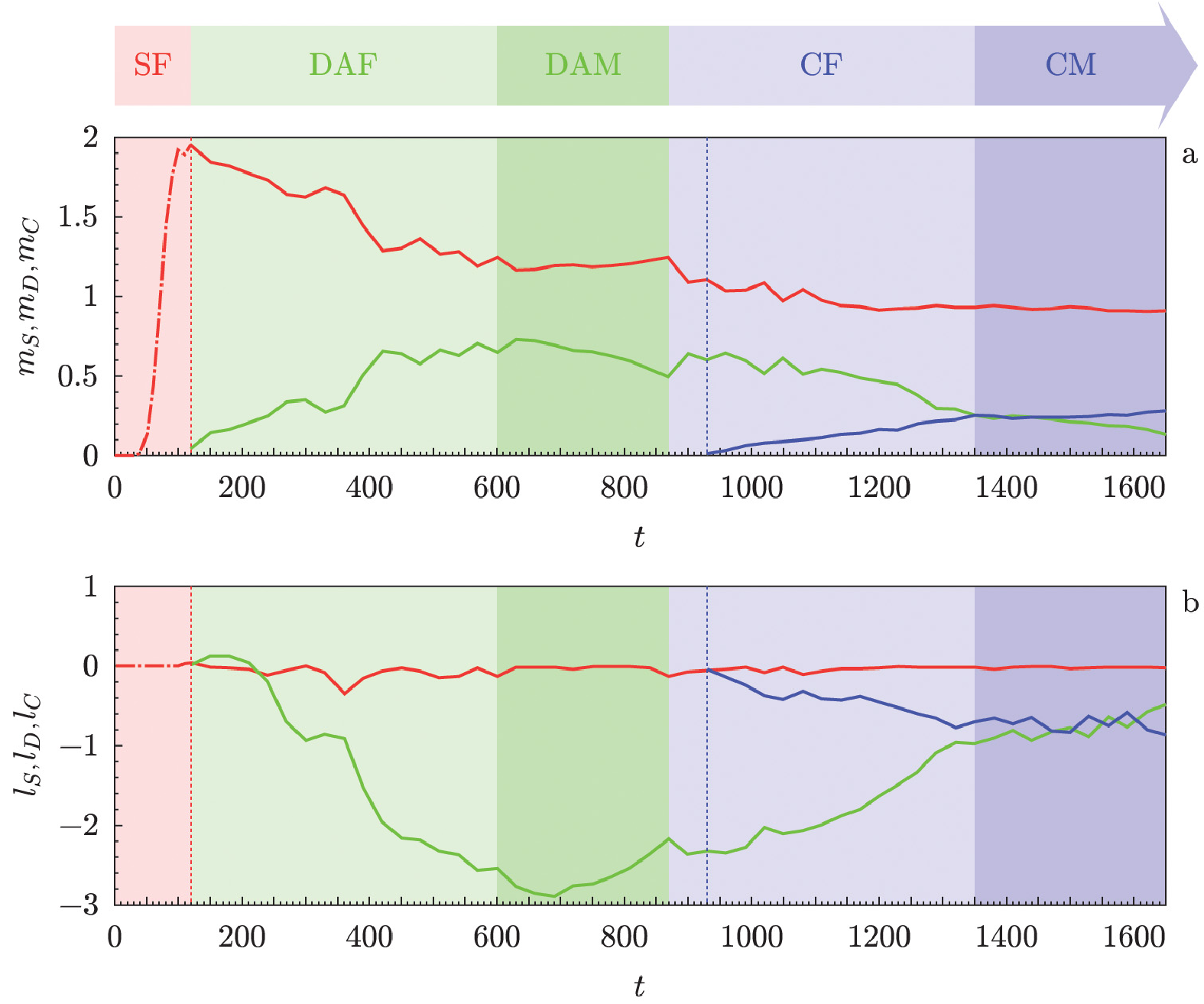}
\caption{(a) Time series of the mass of the central star, disk, arm, and companion, and (b) the angular momentum of them. The red lines represent the mass $m_{S}$ and angular momentum $l_{S}$ of the central star, the green lines represent the mass $m_{D}$ and angular momentum $l_{D}$ of the disk with spiral arms, and the blue lines represent the mass $m_{C}$ and angular momentum $l_{C}$ of the companion. The red dash-dot lines also represent the mass and momentum of gas clumps forming the central star. The red lines show the formation time of the central star ($t=120$), and the blue dash lines show the formation time of the companion ($t=930$).\label{fig:time_series.eps}}
\end{figure}
\unskip

\section{Dynamic formation of a star, a disk, arms, and companions}
\label{Dynamic formation of a star, a disk, arms, and companions}

Let us take a closer look at a series of formation processes going from a central star to a Keplerian disk, spiral arms, and various stellar, substellar, and planetary companions, which are covered by the Astro CML consistently. Interestingly, all of the formation processes are indeed dynamic, with the central star in the ``starring'' role, and they are performed on the disk gravitational stability (the Toomre $Q$ \cite{Toomre} $>1$) but the star gravitational instability (the Toomre $Q<1$). Once the spiral arms are matured around the central star in ``great supporting'' roles, the collaboration of both performers causes a dramatic evolutionary event called ``arm-crossing companion formation,'' and a series of formation processes reaches its climax.

Figure~\ref{fig:ss_30stp_alpha_05(log).eps} shows the series of formation processes with snapshots of the gas clump mass $m_{ij}^{t}$ at $t=0,30,\cdots,1650$ for $\alpha=0.05$  (you can also watch the simulation video \cite{Nozawa10}). It took only 1 min to obtain these results on a personal computer due to fast computation in CML advantages \cite{Kaneko,Kanekos,Kanekob}. Figure~\ref{fig:time_series.eps} shows a classification of the series of formation processes in Figure~\ref{fig:ss_30stp_alpha_05(log).eps} into five distinctive evolutionary stages using the time series of the mass of the central star, disk, arms, and companion and the angular momentum of them around the central star as follows: central star formation (SF) stage, disk and spiral arm formation (DAF) stage, disk and spiral arm maturation (DAM) stage, companion formation (CF) stage, and companion maturation (CM) stage. Note that the time $t$ used in Sections \ref{Central star formation}-\ref{Overview of multiple companion formation} for explanation is the time of the snapshot, not the exact time of the target event. Hereafter, given that red dwarfs are the most common stars in the Milky Way \cite{Henry}, we evaluate the companion mass by regarding the mass of the formed central star to be the typical mass of a red dwarf (about 10\% of the solar mass \cite{Parsons}). Note that, in the evaluation with the central star mass regarded to be the maximum mass of a red dwarf (about 50 \% of the solar mass), companions in the stellar mass range also appear.

\subsection{Central star formation (SF)}
\label{Central star formation}

The central star formation stage is shown in the snapshots at $t=0$, $30$, $\cdots$, $120$ in Figure~\ref{fig:ss_30stp_alpha_05(log).eps} and in the time series at the SF stage on the light red area in Figure~\ref{fig:time_series.eps}. Gas clumps transport their mass (the red dash-dot line in Figure~\ref{fig:time_series.eps}a, increasing from 0\% to 98\% of the initial total mass) and a tiny angular momentum (the red dash-dot line in Figure~\ref{fig:time_series.eps}b, $-3\times10^{-4}$ in clockwise rotation) generated by viscoelastic advection to the central part of the lattice, as they repeatedly fall and deform due to their gravitational interaction and viscoelastic advection. Around $t=120$ (the red dash line in Figure~\ref{fig:ss_30stp_alpha_05(log).eps}), a central star consisting of four massive gas clumps is formed at the lattice center, lattice points 24~24, 25~24, 24~25, and 25~25, whose mass $m_{S}$ is almost the same as the initial total mass of 2 (98\% of the initial total mass). The remaining tiny gas clumps (2\% of the initial total mass) approach the central star closely, rotate, and move away from it, like spacecraft flybys, which provides an opposite angular momentum $l_{S}$ for the central star. Note that the central star is stably formed through rapid relaxation, even if it starts from either such a nonstationary initial state ($\myvec{v}_{ij}^{t}=\myvec{0}$) or a more stationary initial state with an angular momentum ($\myvec{v}_{ij}^{t}\neq\myvec{0}$, data not shown).

\subsection{Disk and spiral arm formation (DAF)}
\label{Disk and spiral arm formation}

The disk and spiral arm formation stage is shown in the snapshots at $t=120$, $150$, $\cdots$, $600$ in Figure~\ref{fig:ss_30stp_alpha_05(log).eps} and in the time series at the DAF stage on the light green area in Figure~\ref{fig:time_series.eps}. The formed central star repeats high-dimensional chaotic gas ejections that transport its mass $m_{S}$ (the red line in Figure~\ref{fig:time_series.eps}a, oscillating and decreasing from 98\% to 60\% of the initial total mass) and angular momentum $l_{S}$ (the red line in Figure~\ref{fig:time_series.eps}b, $-8\times10^{-2}$ on average in clockwise rotation) to its surroundings due to its gravitational instability (the Toomre $Q<1$). This leads to the formation and growth of a Keplerian disk consisting of Keplerian gas clumps and spiral arms consisting of jammed Keplerian gas clumps in gravitational stability (the Toomre $Q>1$) with mass $m_{D}$ (the green line in Figure~\ref{fig:time_series.eps}a, oscillating and increasing from 2\% to 40\% of the initial total mass) and angular momentum $l_{D}$ (the green line in Figure~\ref{fig:time_series.eps}b, increasing from $1\times10^{-2}$ to $-3$ in clockwise rotation). The mass range of the Keplerian disk and spiral arms, ranging from 10\% to 60\% of the central star mass $m_{S}$, is in good agreement with the observed facts, including the small and light ones, which are essential for the earliest stage of protoplanetary systems \cite{Okoda}.

\subsection{Disk and spiral arm maturation (DAM)}
\label{Disk and spiral arm maturation}

The disk and spiral arm maturation stage is shown in the snapshots at $t=600$, $630$, $\cdots$, $870$ in Figure~\ref{fig:ss_30stp_alpha_05(log).eps} and in the time series at the DAM stage on the dark green area in Figure~\ref{fig:time_series.eps}. Once the Keplerian disk and spiral arms have grown sufficiently, gas ejection from the central star reaches a lull. The central star holds constant mass $m_{S}$ (the red line in Figure~\ref{fig:time_series.eps}a, keeping about 60\% of the initial total mass) and angular momentum $l_{S}$ (the red line in Figure~\ref{fig:time_series.eps}b, $-4\times10^{-2}$ in clockwise rotation), and thus the disk and arms also maintain almost constant mass $m_{D}$ (the green line in Figure~\ref{fig:time_series.eps}a, keeping around 40\% of the initial total mass) and angular momentum $l_{D}$ (the green line in Figure~\ref{fig:time_series.eps}b, $-3$ in clockwise rotation), under Keplerian rotation, except for small amounts out through the lattice boundary. This leads to the maturation of spiral arms to balanced grand design ones after $t=720$.

\subsection{Companion formation (CF)}
\label{Companion formation}

The companion formation stage is shown in the snapshots at $t=870$, $900$, $\cdots$, $1350$ in Figure~\ref{fig:ss_30stp_alpha_05(log).eps} and in the time series at the CF stage on the light blue area in Figure~\ref{fig:time_series.eps}. Once the angular momentum change of the central star becomes dominant due to the mature spiral arm retaining mass ($\sim 2\times10^{-3}$ on the 4-neighborhood average) and angular momentum ($\sim -2\times10^{-5}$ on the 4-neighborhood average) by jammed Keplerian gas \cite{Nozawa2}, the contracting central star ejects gas again from $t=870$ to $900$, like a skater spinning faster by drawing arms, under the conservation of angular momentum \cite{Nozawa}. The ejected gas becomes a new spiral arm that transports the central star mass $m_{S}$ (the red line in Figure~\ref{fig:time_series.eps}a, oscillating and decreasing from 60\% to 50\% of the initial total mass) and angular momentum $l_{S}$ (the red line in Figure~\ref{fig:time_series.eps}b, decreasing from $-1\times10^{-1}$ to $-7\times10^{-2}$ in clockwise rotation), which crosses with the mature spiral arm around $t=930$ (the blue dash line in Figure~\ref{fig:time_series.eps}). This sudden increase in mass (5 times from 0.2\% to 1\% of $m_{S}$) and angular momentum (10 times from $-2\times10^{-5}$ to $-2\times10^{-4}$) at the intersection located at lattice points 27~32, 28~32, 27~33, and 28~33, leads to the formation of a companion in the planet mass range, consisting of four rotating dense Keplerian gas clumps, without the radial drift barrier. We will call this dramatic evolutionary event ``arm-crossing companion formation,'' which differs significantly from previous companion formation scenarios, such as core accretion \cite{Hayashi,Safronov} and disk instability \cite{Cameron,Kuiper}. We would suggest that the ``arm-crossing companion formation'' scenario could bring us to the origin of hot Jupiters, which is at issue in these companion formation scenarios \cite{Dawson}. This is because the mature spiral arm, which consists of gas clumps with dust and retains mass and angular momentum, is temperature-insensitive (so that it does not require a ``snow line'' argument) and can produce both rocky and gaseous companions.

From $t=930$ to $1350$, the formed companion star grows up stably, while increasing its mass $m_{C}$ (the blue line in Figure~\ref{fig:time_series.eps}a, smoothly increasing from 1\% to 30\% of $m_{S}$, from a planet to a brown dwarf) and angular momentum $l_{C}$ (the blue line in Figure~\ref{fig:time_series.eps}b, increasing from $-4\times10^{-2}$ to $-7\times10^{-1}$ in clockwise rotation) by absorbing the mass $m_{D}$ (the green line in Figure~\ref{fig:time_series.eps}a, oscillating and decreasing from 50\% to 30\% of $m_{S}$) and angular momentum $l_{D}$ (the green line in Figure~\ref{fig:time_series.eps}b, decreasing from $-2$ to $-1$ in clockwise rotation) from the mature disk. The companion also increases its rotational angular momentum 100 times from $-2\times10^{-4}$ to $-2\times10^{-2}$, and its constituent gas clumps become like a pressure bump \cite{Marel} that traps and moves together dust particles, which prevents the radial drift of the dust particles.

\subsection{Companion maturation (CM)}
\label{Companion maturation}

\begin{figure}[t]
\isPreprints{}{
\begin{adjustwidth}{-\extralength}{0cm}
\centering
}
\includegraphics[scale=0.93]{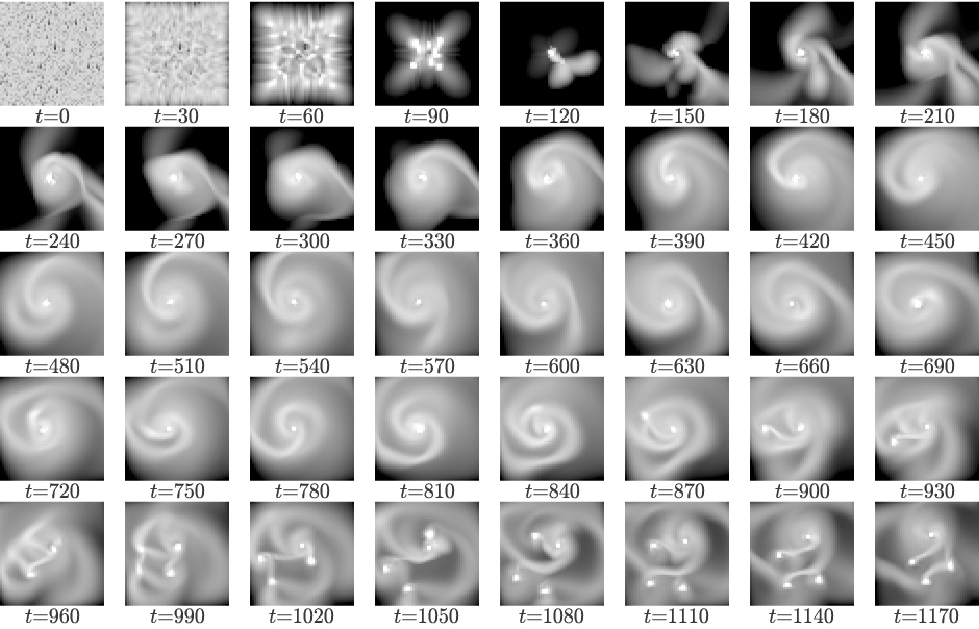}
\isPreprints{}{
\end{adjustwidth}
}
\caption{A series of formation processes at $\alpha=0.06$ from a central star to a Keplerian disk, spiral arms, and three companions. In each snapshot at $t=0$, $30$, $\cdots$, $1170$, the logarithm of gas clump masses, $\log_{10}m_{ij}^{t}$ ($i=0,1,\cdots,49$ and $j=0,1,\cdots,49$), is plotted in grayscale in the range of $1\times 10^{-8}$ to $1\times 10^{-2}$.\label{fig:ss_30stp_alpha_06(log).eps}}
\end{figure}

The companion maturation stage is shown in the snapshots at $t=1350$, $1380$, $\cdots$, $1650$ in Figure~\ref{fig:ss_30stp_alpha_05(log).eps} and in the time series at the CM stage on the dark blue area in Figure~\ref{fig:time_series.eps}. Once the companion has grown sufficiently, it stops absorbing the disk mass $m_{D}$ (the green line in Figure~\ref{fig:time_series.eps}a, keeping around 30\% of $m_{S}$), and thus its mass $m_{C}$ becomes constant (the blue line in Figure~\ref{fig:time_series.eps}a, keeping about 30\% of $m_{S}$, a brown dwarf). Interestingly, the mass ratio of the formed companion to the central star is consistent with the most typically observed mass ratio \cite{Duquennoy,Chauvin}. On the other hand, its angular momentum $l_{C}$ (the blue line in Figure~\ref{fig:time_series.eps}b, oscillating in reverse phase with $l_{D}$ around $-7\times10^{-1}$ on average in clockwise rotation) is violently exchanged with the disk angular momentum $l_{D}$ (the green line in Figure~\ref{fig:time_series.eps}b, oscillating in reverse phase with $l_{C}$ around $-8\times10^{-1}$ on average in clockwise rotation) due to the companion and disk colliding elastically. This forms spiral arms extending from the companion, which enables the ``arm-crossing companion formation'' further away from the central star (e.g., see the snapshots in Figure~\ref{fig:ss_30stp_alpha_06(log).eps}). Additionally, from $t=1470$ to $1650$, the disk mass $m_{D}$ is isotropically swept out (the green line in Figure~\ref{fig:time_series.eps}a, decreasing from 30\% to 10\% of $m_{S}$) by a violent collision with the Keplerian rotating companion, and the disk is now occupied only by the spiral arms and stars. Although we did not calculate it in the present simulation due to limitations in the lattice size, we suspect that this swept-out gas may be the origin of the ring structure observed far from the central star \cite{Cieza}. If this is the case, the number of rings would be expected to correspond to that of companions that exist much farther inward than the rings.

We would emphasize that the diversity of evolutionary stages described above emerged spontaneously as a result of a series of evolutions in the collective behavior of gas clumps under the universality of chaotic dynamics due solely to the gravitational interaction and viscoelastic advection of gas clumps containing small amounts of dust. Needless to say, this consistent scenario for the evolution from a star to a disk, arms, and companions is based on insightful suggestions from the Astro CML.

\subsection{Overview of multiple companion formation}
\label{Overview of multiple companion formation}

Even with a larger mixing coefficient $\alpha$ of $0.06$, the series of formation processes go through the same five evolutionary stages, but it differs in that multiple companions are formed in the companion formation stage. Now, let us overview this series of formation processes. Figure~\ref{fig:ss_30stp_alpha_06(log).eps} shows the series of formation processes with snapshots of the gas clump mass $m_{ij}^{t}$ at $t=0,30,\cdots,1170$ for $\alpha=0.06$ (you can also watch the simulation video \cite{Nozawa10}).

The central star formation stage goes from $t=0$ to $120$, the disk and spiral arm formation stage from $t=120$ to $480$, and the disk and spiral arm maturation stage from $t=480$ to $780$. Three diverse companions form successively after this stage. We can summarize them by companion as follows: For the first companion, the formation stage goes from $t=780$ to $870$ (8\% of $m_{S}$, a planet) and the maturation stage from $t=870$ to $1170$ (25\% of $m_{S}$, a brown dwarf); for the second companion, the formation stage goes from $t=870$ to $990$ (3\% of $m_{S}$, a planet) and the maturation stage from $t=990$ to $1170$ (8\% of $m_{S}$, a planet); and for the third companion, the formation stage goes from $t=900$ to $990$ (6\% of $m_{S}$, a planet) and the maturation stage from $t=990$ to $1170$ (10\% of $m_{S}$, a planet).

The first companion, which is formed around $t=840$, begins to extend its spiral arms around $t=870$, which are given by the logarithmic spiral. After that, gas ejection from the central star occurs at equal intervals of azimuthal angle around $t=900$ and $t=930$ due to the gravitational interaction (tidal deformation) between the first companion and the central star, and then a second companion is born around $t=930$ and a third around $t=960$ successively on the logarithmic spiral extending from the first companion. Furthermore, as can be seen from the snapshots at $t=990$, $1020$, and $1050$, the spiral arm extending from the first companion lifts the second companion, and that from the second companion lifts the third companion, even moving them away from the central star. Based on these results, we are confident that the ``arm-crossing companion formation'' may provide the inevitable of the Titius-Bode law \cite{Kanekop}, which was previously regarded as just a coincidence. More details will be reported soon.

\section{Summary and discussion}
\label{Summary and discussion}

In this paper, we have reported the Astro CML, which consistently covers a series of formation processes of four types of astronomical objects, whose existence has been confirmed in protoplanetary systems, including a central star, a Keplerian disk, spiral arms, and various stellar, substellar, and planetary companions. The Astro CML simulated a series of evolutionary stages in the collective behavior of discrete gas clumps containing small amounts of dust, based on the viscoelastic and chaotic dynamics consisting of a minimal set of an Eulerian procedure for the flow formation of gas clumps due to gravitational interaction and a Lagrangian procedure for the collision and mixture of gas clumps due to viscoelastic advection. In the simulations, all four types of astronomical objects in protoplanetary systems were successfully reproduced. All of each object is formed not based on either the conventional core accretion or disk instability, but on the star instability, and experiences indeed a dynamic formation process, with the central star in the ``starring'' role and the spiral arms in the ``great supporting'' role. Notably, the ``arm-crossing companion formation'' that causes a sudden increase in mass and angular momentum at the arm intersection enables the formation of companions in the planet or brown dwarf or star mass range without experiencing the radial drift.

The formation process of each object was classified into five evolutionary stages: (1)~Central star formation stage: Gas clumps transport their mass and angular momentum to the forming central star as they repeatedly fall and deform due to gravitational interaction and viscoelastic advection. (2)~Disk and spiral arm formation stage: The formed central star repeats high-dimensional chaotic gas ejections due to gravitational instability, thus leading to the formation and growth of a Keplerian disk and spiral arms. (3)~Disk and spiral arm maturation stage: Once the Keplerian disk and spiral arms have grown sufficiently, gas ejection from the central star settles down, resulting in the maturation of spiral arms to a balanced grand design. (4)~Companion formation stage: Gas ejection from the central star forms a new spiral arm, which crosses with the mature spiral arm, thus leading to the formation and growth of a stellar, substellar, or planetary companion. (5)~Companion maturation stage: Once the companion has grown enough to stop absorbing the disk mass, it begins to exchange angular momentum with the disk, and thus spiral arms extend from the companion, leading to the ``arm-crossing companion formation'' even at positions far from the central star. This ``arm-crossing companion formation'' can repeat to form a series of companions that move away from the central star one after the other.

A series of formation processes of astronomical objects in protoplanetary systems with a central star as ``starring,'' which the Astro CML has demonstrated, is considered to be closely related to CI in high-dimensional chaotic dynamical systems, in particular, to astronomical CI \cite{Konishi}. In fact, the nonlinear dynamics of gas clump velocity inside the central star, whose two massive gas clump elements interact gravitationally, is expressed as a map similar to the logistic map with its mass as a bifurcation parameter \cite{Nozawa7,Nozawa8}. This map does not follow the Feigenbaum scenario and exhibits a bifurcation phenomenon to high-dimensional space, i.e., high-dimensional chaotic gas ejection, due to the breaking of the low-dimensional constraints in the middle of a period-doubling bifurcation \cite{Nozawa9}. This bifurcation phenomenon will appear in systems with abundant spatial degrees of freedom, such as the universe. In the Astro CML, the central star is not described as a singular point but as identical to the other gas clumps except for its large mass, suggesting that similar maps should underlie any collective behavior of gas clumps. The results of our investigation into astronomical CI \cite{Konishi,Nozawa9}, such as observed in Mira variables \cite{Yanagitaa}, by regarding the Astro CML as a coupled map to which such high-dimensional bifurcation is intrinsic, will be reported in future papers.
\vspace{6pt} 

\funding{This research received no external funding.}

\dataavailability{The original contributions presented in this study are included in the article. Further inquiries can be directed to the corresponding author.} 

\acknowledgments{The author would like to thank Professor Tetsuo Deguchi for his valuable comments and warm encouragement. The author would also like to thank Professor Kunihiko Kaneko for his extensive discussions and insightful suggestions on simulation results. Finally, the author is also grateful to the members of the Universal Biology group at the Niels Bohr Institute for their helpful feedback.}

\conflictsofinterest{The authors declare no conflicts of interest.} 

\clearpage

\appendixtitles{yes}
\appendixstart
\appendix
\section[\appendixname~\thesection]{Viscoelastic advection}
\label{Viscoelastic advection}

The ``viscoelasticity'' in the viscoelastic advection procedure $T_{a}$ is succinctly represented by the displacement of a gas clump containing a small amount of dust, which appears in Equation (\ref{eqn:tr_kl}). We now derive this displacement based on a difference in velocity relaxation arising between the viscous behavior of light gas particles and the elastic behavior of heavy dust particles in the flow, and discuss the small angular momentum generated by the viscoelastic advection using the obtained displacement.

We consider a gas clump of mass $m_{ij}^{t}$ containing a small amount of dust that relaxes over a time $\tau_{a}$ from a velocity $\myvec{v}_{ij}^{t}$ to the velocity $\myvec{v}_{ij}^{*}$ resulting from the gravitational interaction procedure $T_{g}$. The momentum change of the gas clump is given by
\begin{align}
\label{eqn:mcgc}
m_{ij}^{t}\left(\myvec{v}_{ij}^{*}-\myvec{v}_{ij}^{t}\right)=\frac{\left(1-\alpha\right)m_{ij}^{t}\left(\myvec{v}_{ij}^{*}-\myvec{v}_{ij}^{t}\right)}{\alpha\tau_{a}}\alpha\tau_{a}+\frac{\alpha m_{ij}^{t}\left(\myvec{v}_{ij}^{*}-\myvec{v}_{ij}^{t}\right)}{\left(1-\alpha\right)\tau_{a}}\left(1-\alpha\right)\tau_{a}.
\end{align}
The first and second terms in Equation (\ref{eqn:mcgc}) represent the fast relaxation of light gas particles over a short time of $\alpha\tau_{a}$ and slow relaxation of heavy dust particles over a long time of $(1-\alpha)\tau_{a}$ from the flow $\myvec{v}_{ij}^{t}$ to the flow $\myvec{v}_{ij}^{*}$ inside the gas clump, respectively.

\begin{figure}[!b]
\isPreprints{\centering}{}
\includegraphics[scale=0.5]{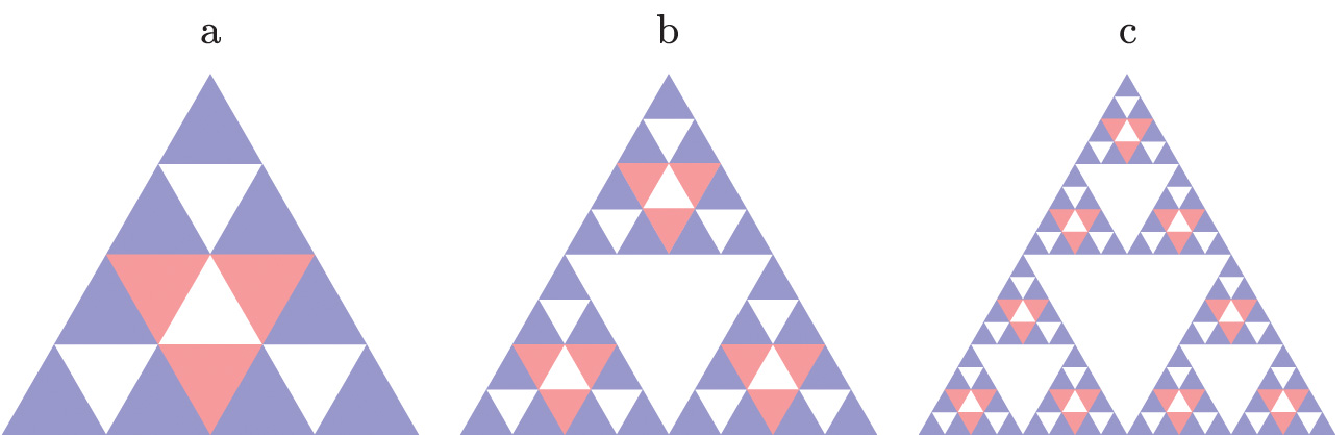}
\caption{Schematic illustration of a gas clump consisting of clad particles (i.e., gas-clad dust particles). (a)~$N_{d}=1$ ($N_{g}=9$); (b)~$N_{d}=3$ ($N_{g}=27$); and (c)~$N_{d}=9$ ($N_{g}=81$), with the mixing coefficient $\alpha=0.25$. In each figure, the blue triangle represents a gas particle, and the group of three red triangles represents a dust particle. \label{fig:gas_clad_dust_particle.eps}}
\end{figure}

These fast and slow relaxations are caused by gas particles of mass $m_{g}$ and a dust particle of mass $m_{d}$ clustering and moving together without phase separation (forming a clad particle, i.e., gas-clad dust particle, such as cosmic dust \cite{Jessberger,Fechtig}) while exchanging their respective momenta according to
\begin{align}
\label{eqn:medg}
\alpha m_{d}\left(\myvec{v}_{ij}^{*}-\myvec{v}_{ij}^{t}\right)=\left(1-\alpha\right)m_{g}\left(\myvec{v}_{ij}^{*}-\myvec{v}_{ij}^{t}\right).
\end{align}
Equation (\ref{eqn:medg}) indicates that the fast relaxation of gas particles is the result of as many as $(1-\alpha)/\alpha$ gas particles exchanging momentum in collision with one dust particle, and the slow relaxation of dust particles is the result of as few as $\alpha/(1-\alpha)$ dust particles exchanging momentum in collision with one gas particle. Here, we have assumed the following self-similarity for mass ratios using the number $N_{g}$ of gas particles and the number $N_{d}$ of dust particles:
\begin{align}
\label{eqn:ssdg}
\frac{m_{d}}{m_{g}}=\frac{N_{g}m_{g}}{N_{d}m_{d}}=\frac{1-\alpha}{\alpha},
\end{align}
where the mass ratio $m_{d}/m_{g}$ of a heavy dust particle to a light gas particle is equal to the mass ratio $N_{g}m_{g}/N_{d}m_{d}$ of the large amount of gas to the small amount of dust in the gas clump. The particle number ratio of gas to dust is then $N_{g}/N_{d}=\{(1-\alpha)/\alpha\}^{2}$. Figure~\ref{fig:gas_clad_dust_particle.eps} schematically represents an example of a gas clump consisting of gas-clad dust particles (gas and dust particles with self-similarity).

\unskip

\begin{figure}[!b]
\isPreprints{\centering}{}
\includegraphics[scale=0.5]{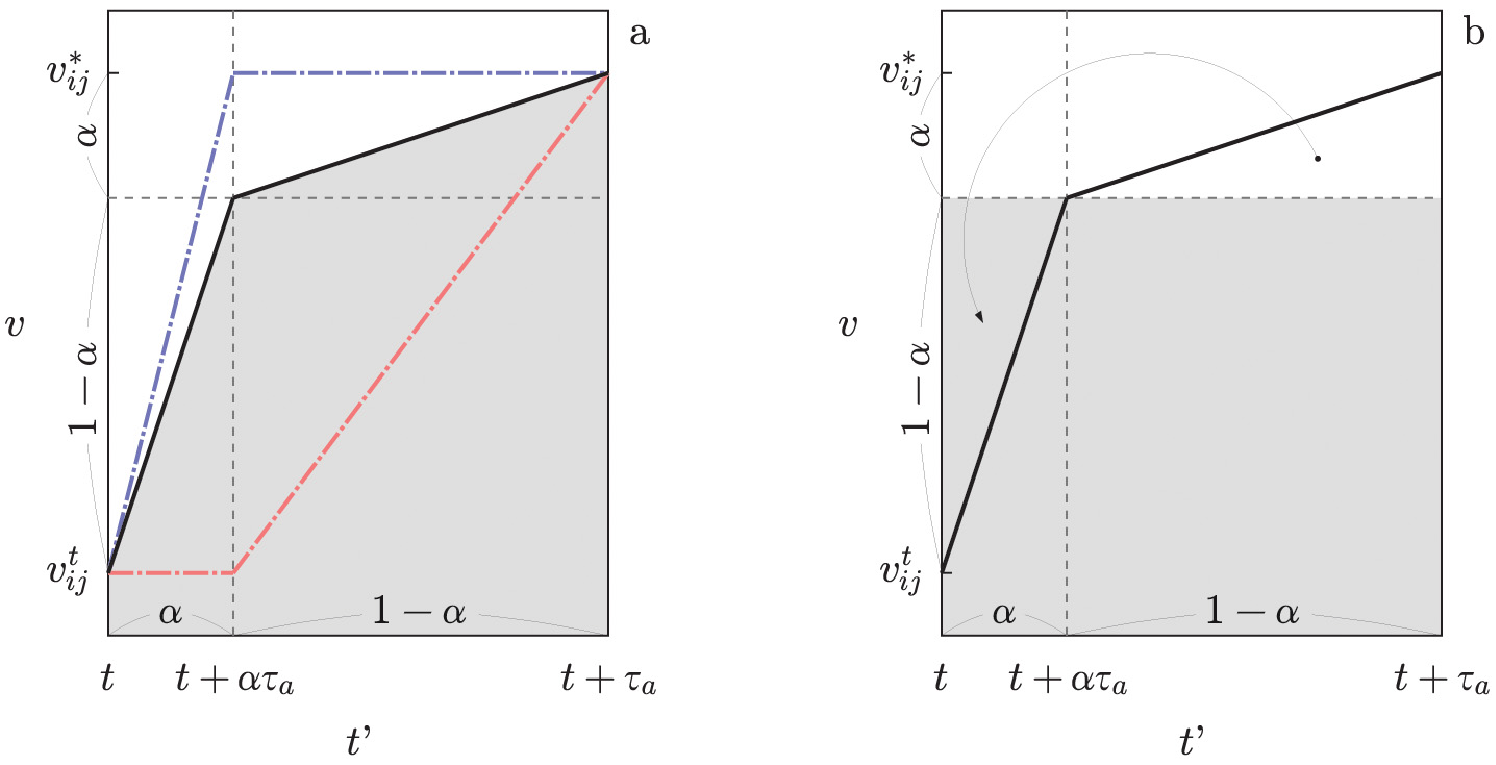}
\caption{(a) Temporal change in the gas clump velocity $\myvec{v}(t')$ with a relaxation difference between gas and dust, and (b) derivation of its displacement. The black solid line represents $\myvec{v}(t')$, and the blue and red dash-dot lines represent the gas and dust velocities, respectively. The gray area in Figure~\ref{fig:relaxation_curve.eps}a shows the displacement $\Delta\myvec{r}_{ij}^{*}$ of the gas clump and has the same size $\{(1-\alpha)\myvec{v}_{ij}^{*}+\alpha\myvec{v}_{ij}^{t}\}\tau_{a}$ as the gray area in Figure~\ref{fig:relaxation_curve.eps}b. This is because the white and gray right triangles are congruent. \label{fig:relaxation_curve.eps}}
\end{figure}

Assuming, for simplicity, linearity with time (uniformly accelerated motion) for fast and slow velocity relaxation in Equation (\ref{eqn:mcgc}), respectively, the velocity $\myvec{v}(t')$ of the gas clump at time $t'$ ($t\le t'\le t+\tau_{a}$) follows
\begin{align}
\label{eqn:vpla}
\myvec{v}(t')=\left\{
\begin{aligned}
&\myvec{v}_{ij}^{t}+\frac{\left(1-\alpha\right)\left(\myvec{v}_{ij}^{*}-\myvec{v}_{ij}^{t}\right)}{\alpha\tau_{a}}\left(t'-t\right),\quad &t\le t'<t+\alpha\tau_{a},\\
&\left(1-\alpha\right)\myvec{v}_{ij}^{*}+\alpha\myvec{v}_{ij}^{t}+\frac{\alpha\left(\myvec{v}_{ij}^{*}-\myvec{v}_{ij}^{t}\right)}{\left(1-\alpha\right)\tau_{a}}\left\{t'-\left(t+\alpha\tau_{a}\right)\right\},\quad &t+\alpha\tau_{a}\le t'\le t+\tau_{a}.
\end{aligned}
\right.
\end{align}
Figure~\ref{fig:relaxation_curve.eps}a shows the temporal change in the gas clump velocity $\myvec{v}(t')$ (the black solid line), plotted along with those in the velocity of gas particles (the blue dash-dot line) and velocity of dust particles (the red dash-dot line) inside the gas clump. From time $t$ to $t+\alpha\tau_{a}$, the gas particles exhibit viscous behavior and their velocity quickly relaxes into the flow $\myvec{v}_{ij}^{*}$, whereas the dust particles exhibit elastic behavior and their velocity remains constant at $\myvec{v}_{ij}^{t}$. After the gas velocity has relaxed at time $t+\alpha\tau_{a}$, the dust particles begin to exhibit viscous behavior, and from time $t+\alpha\tau_{a}$ to $t+\tau_{a}$, their velocity slowly relaxes into the flow $\myvec{v}_{ij}^{*}$. As a result, gas clumps containing small amounts of dust will exhibit viscoelastic velocity relaxation. We thus call such advection with this relaxation ``viscoelastic advection.'' From Figure~\ref{fig:relaxation_curve.eps}b, the displacement $\Delta\myvec{r}_{ij}^{*}$ of the gas clump upon viscoelastic advection (the size of the gray area) becomes
\begin{align}
\label{eqn:drij}
\Delta\myvec{r}_{ij}^{*}=\left\{\left(1-\alpha\right)\myvec{v}_{ij}^{*}+\alpha\myvec{v}_{ij}^{t}\right\}\tau_{a},
\end{align}
which yields $\tilde{\myvec{r}}_{ij}^{*}=\myvec{r}_{ij}+\Delta\myvec{r}_{ij}^{*}$ of Equation (\ref{eqn:tr_kl}) using Equation (\ref{eqn:Tg}).

\unskip

Note that the velocity $\myvec{v}(t')$ of the gas clump can also be obtained in another way \cite{Nozawa5}. During velocity relaxation, a gas particle with velocity $\myvec{v}_{g}$ and a dust particle with velocity $\myvec{v}_{d}$ follow
\begin{align}
\label{eq:devr}
\left\{
\begin{aligned}
\frac{d\myvec{v}_{g}}{dt'}&=-\frac{1}{\alpha\tau_{a}}(\myvec{v}_{g}-\myvec{v}_{ij}^{*}),\\
\frac{d\myvec{v}_{d}}{dt'}&=-\frac{1}{\tau_{a}}(\myvec{v}_{d}-\myvec{v}_{g}).
\end{aligned}
\right.
\end{align}
The velocity $\myvec{v}_{ij}^{*}$, which appears in the gas velocity relaxation, is based on the aforementioned elastic behavior of the dust particles. Solving Equation (\ref{eq:devr}) results in
\begin{align}
\myvec{v}_{g}\left(t'\right)&=\myvec{v}_{ij}^{*}-\left(\myvec{v}_{ij}^{*}-\myvec{v}_{ij}^{t}\right)\mathrm{e}^{-\frac{t'-t}{\alpha\tau_{a}}},\\
\myvec{v}_{d}\left(t'\right)&=\myvec{v}_{ij}^{*}+\frac{\alpha}{1-\alpha}\left(\myvec{v}_{ij}^{*}-\myvec{v}_{ij}^{t}\right)\mathrm{e}^{-\frac{t'-t}{\alpha\tau_{a}}}-\frac{1}{1-\alpha}\left(\myvec{v}_{ij}^{*}-\myvec{v}_{ij}^{t}\right)\mathrm{e}^{-\frac{t'-t}{\tau_{a}}},
\end{align}
and then approximating them piecewise linearly, such as with $\exp(-(t'-t)/a)\sim 1-(t'-t)/a$ ($t\le t'\le t+a$), yields Equation (\ref{eqn:vpla}).

Comparing the angular momenta $\myvec{l}(t+\tau_{a};\alpha)$ of a gas clump consisting only of gas ($\alpha=0$) and a gas clump containing a small amount of dust ($\alpha\ll1$), using the displacement $\Delta\myvec{r}_{ij}^{*}$ due to viscoelastic advection of Equation (\ref{eqn:drij}), we obtain
\begin{align}
\label{eqn:dl}
\myvec{l}(t+\tau_{a};\alpha)-\myvec{l}(t+\tau_{a};0)&=\left[\myvec{r}_{ij}+\left\{\left(1-\alpha\right)\myvec{v}_{ij}^{*}+\alpha\myvec{v}_{ij}^{t}\right\}\tau_{a}\right]\times m_{ij}^{t}\myvec{v}_{ij}^{*}
\nonumber\\
&-\left(\myvec{r}_{ij}+\myvec{v}_{ij}^{*}\tau_{a}\right)\times m_{ij}^{t}\myvec{v}_{ij}^{*}
\nonumber\\
&=\alpha\tau_{a}\myvec{v}_{ij}^{t}\times m_{ij}^{t}\myvec{v}_{ij}^{*}.
\end{align}
Equation (\ref{eqn:dl}) shows that viscoelastic advection produces a small angular momentum due to the elastic behavior $\alpha\tau_{a}\myvec{v}_{ij}^{t}$ of dust particles (the flat red dash-dot line from time $t$ to $t+\alpha\tau_{a}$ in Figure~\ref{fig:relaxation_curve.eps}a). This angular momentum not only leads the central and companion stars to rotate, but also transports mass and angular momentum through the rotation.

\isPreprints{}{
\begin{adjustwidth}{-\extralength}{0cm}
}

\reftitle{References}

\end{document}